\documentclass[iop,apj,numberedappendix]{emulateapj}

\usepackage{epstopdf}
\usepackage{natbib}
\usepackage{graphicx}
\usepackage{amsmath,amsfonts}
\slugcomment{Accepted for publication in ApJ, August 23, 2012}

\def\gtwid{\mathrel{\raise.3ex\hbox{$>$\kern-.75em\lower1ex\hbox{$\sim$}}}}
\def\ltwid{\mathrel{\raise.3ex\hbox{$<$\kern-.75em\lower1ex\hbox{$\sim$}}}}

\defcitealias{IPDF}{JG12}
\setcitestyle{notesep={; }}

\begin{document}
\title{Constraining the Vela Pulsar's Radio Emission Region Using Nyquist-Limited Scintillation Statistics}
\shorttitle{Vela Pulsar Emission Size}

\author{M. D. Johnson, C. R. Gwinn}
\affil{Department of Physics, University of California, Santa Barbara, California 93106, USA}
\email{ michaeltdh@physics.ucsb.edu,cgwinn@physics.ucsb.edu} 

\author{P. Demorest}
\affil{National Radio Astronomy Observatory, 520 Edgemont Road, Charlottesville, VA 22093 USA}
\email{pdemores@nrao.edu}

\shortauthors{Johnson et al.}


\begin{abstract}
Using a novel technique, we achieve ${\sim}$100 picoarcsecond resolution and set an upper bound of less than 4 km for the characteristic size of the Vela pulsar's emission region. Specifically, we analyze flux-density statistics of the Vela pulsar at 760 MHz. Because the pulsar exhibits strong diffractive scintillation, these statistics convey information about the spatial extent of the radio emission region. We measure both a characteristic size of the emission region and the emission sizes for individual pulses. Our results imply that the radio emission altitude for the Vela pulsar at this frequency is less than 340 km.
\end{abstract}


\section{Introduction}
\label{sec::Intro}
\subsection{Pulsar Emission and Scintillation}

Owing to their extraordinary emission, pulsars are invaluable probes of a vast array of extreme physics, from the supranuclear density interiors and the ${\sim}10^{12}\mathrm{\ G}$ magnetospheres of neutron stars, to the turbulent plasma of the interstellar medium (ISM). 
The regularity of their time-averaged emission has been applied to sensitive tests of general relativity \citep{Taylor_Weisberg_89,Kramer_GR}, and the achieved timing precision holds promise for direct observation of gravitational waves \citep{Detweiler_GW,Jenet_GW,IPTA}. Yet, remarkably, no consensus exists for the precise origin of this emission; its geometry, stability, and physical mechanism remain enigmatic, even after nearly a half-century of study \citep{Melrose_2004,Verbiest_2009}. The incredible compactness of the pulsar and its surrounding plasma prohibit traditional imaging of the emission process, but a number of methods can achieve the requisite resolution, using the interstellar plasma as an effective `lens' \citep{Lovelace_Thesis,Backer_75,Cordes83,CornwellNarayan93,ISO,Shishov_2010}. 

Variations of this stochastic lens give rise to the observed scintillation, and statistics of these variations convey information about the spatial structure of the emission. 
Longer observing wavelengths enhance the signature of this structure because the stronger scattering effectively enlarges the aperture. However, the increased spatial resolution comes at the expense of spectral resolution, and it becomes difficult to decouple the scintillation from source noise and variability. Indeed, the pulse-broadening timescale can easily exceed that of intrinsic pulsar variation.

We can circumvent these difficulties by constructing spectra over accumulation times that include \emph{all} pulsed power. Intrinsic sub-pulse variability then affects the correlation of spectral noise but not the mean spectrum or single-channel statistics \citep{Intermittent_Noiselike_Emission}. 
Furthermore, we can distinguish the effects of extended source geometry from the scattering evolution and pulsar variability by varying the degree of temporal averaging; with no averaging, only the source emission structure contributes. Additionally, we can identify the emission sizes of individual pulses. 
We have presented a detailed mathematical treatment separately \citep[hereafter JG12]{IPDF}.

We constructed spectra in this way and then compared histograms formed from the spectral samples with appropriate models for a strongly scintillating source. 
Such models benefit from minimal assumptions about the nature and distribution of both the scattering material and the underlying emission; this generality is achieved at the expense of resolution in pulse phase. The inferred emission size is therefore a characteristic size of the emission region, quantified via a dimensionless parameter $\gamma_{\mathrm{s}}$, which depends on the spatial standard deviation of source intensity, weighted by integrated flux density. Translating $\gamma_{\mathrm{s}}$ to physical units at the pulsar additionally requires specification of the scattering geometry.

\subsection{Comparison with Previous Work}
\label{sec::Previous}
Several previous investigations have also quantified scintillation statistics in order to estimate the size of the Vela pulsar's emission region. 
In general, for a strongly scattered point source, the scintillation acts upon intensity as a multiplicative gain $\mathcal{G}$ that is the squared modulus of a complex random walk. The probability density function (PDF) of $\mathcal{G}$ is therefore exponential, $P(\mathcal{G}) = e^{-\mathcal{G}}$, with an associated modulation index of unity: $m^2 \equiv \left\langle \mathcal{G}^2 \right \rangle / \left\langle \mathcal{G} \right \rangle^2 - 1 = 1$. 
A spatially-extended emission region smoothes this variation, and so the modified $P(\mathcal{G})$ or reduced modulation index can be used to estimate the emission size \citep{Salpeter67,Cohen67,ISO}.

The principal difficulty in detecting these modifications is the additional noise that arises from the noiselike nature of the source, known as \emph{self-noise}, and from the background. Pulsars are particularly challenging targets because their dramatic intrapulse and interpulse variations also contribute modulation. All of these sources of noise can easily imitate or mask the signature of an extended emission region. Previous efforts have attempted to minimize these sources of noise by spectral averaging in both frequency and time. In contrast, the present technique fully accounts for the effects of self-noise and intrinsic pulsar variability.

For example, \citet{Gwinn97} sought to detect the modification of $P(\mathcal{G})$ through the PDF of the correlation function amplitude on the Tidbinbilla-Parkes baseline at 2.3 GHz. Because this baseline was much shorter than the scale of the diffraction pattern, the observational setup was effectively a zero-baseline interferometer. Their best-fitting model corresponded to an emission region with a FWHM of $460 \pm 110$ km. 
\citet{Gwinn00} then extended this analysis to three gates across the pulse, also at 2.3 GHz. 
They improved the model PDF and quantified the effects of their averaging in frequency and time. They estimated a declining size of the emission region from $440\pm 90\mathrm{\ km}$ to less than $200 \mathrm{\ km}$ across the pulse.

Rather than analyzing distribution functions, \citet{Macquart00} used the modulation index to quantify the effects of emission size for single-dish data. They measured $m^2 \approx 0.9$ at 660 MHz and thereby inferred the extent of source emission to be no more than 50 km ($\mathrm{FWHM}=120$ km). 
Including the contribution of self-noise leads to an inferred FWHM of the emission region of ${\sim}200\mathrm{\ km}$ \citepalias{IPDF}.

More recently, \citet{Vela_18cm} analyzed visibilities on the Tidbinbilla-Mopra baseline at 1.65 GHz and fit models to the projections of the real part and imaginary variance. 
They found the size of the emission region to be large both early and late in the pulse (${\sim}400\mathrm{\ km}$ and ${\sim}800\mathrm{\ km}$, respectively) but near zero in the central portion, where the pulse is strong.

An analysis at lower frequencies has the advantage of increasing the signature of size but the disadvantage of incurring more mixing between the scintillation and the pulsar noise. 
We demonstrate that this difficulty is not prohibitive, and that a careful statistical description of observed power spectra at low frequencies can yield fresh insight into the physics of both interstellar scattering and pulsar emission.

\subsection{Outline of Paper}

In \S\ref{sec::Theory}, we briefly review the theoretical descriptions of pulsar emission and interstellar scintillation. 
Then, in \S\ref{sec::Obs}, we describe our observations and the subsequent construction of both the observed and model PDFs of intensity; we also present the expected residual structure arising from an extended emission region and outline the model fitting procedure. 
Next, in \S\ref{sec::Analysis}, we give our analysis for various pulse selection cuts and derive our estimates for the size of the emission region. In \S\ref{sec::Errors_and_Bias}, we quantify possible biases and systematic errors in our inferences of the size of the emission region. 
Then, in \S\ref{sec::Interpretation}, we interpret our measurements in terms of beamed, dipolar emission, and derive the corresponding emission altitudes.
Finally, in \S\ref{sec::Summary}, we summarize our results, their impact on our understanding of pulsar magnetospheres, and the prospects for future work.

\section{Theoretical Background}
\label{sec::Theory}

\subsection{Pulsar Radio Emission}

A full understanding of pulsar emission is elusive, but proposed scenarios typically invoke a relativistic electron-positron plasma that flows outward from the pulsar polar caps to the light cylinder \citep{Goldreich_Julian}. 
This outflow then generates radio emission via coherent curvature radiation \citep{Sturrock,Ruderman_Sutherland}. The emission is beamed, ducted, and refracted as it propagates through the upper magnetosphere \citep{Barnard_Arons,Arons_Barnard,Lyutikov_Parikh}. 

The precise altitude at which the emission is generated is uncertain, and various analyses suggest emission heights from a few stellar radii \citep{Arons_Scharlemann} to hundreds of stellar radii \citep{Karastergiou_Johnston} for Vela-like pulsars. In the classification and interpretation of \citet{Rankin_Core}, the Vela pulsar is a core-single ($\mathrm{\textbf{S}}_\mathrm{t}$) pulsar, with emission arising effectively at the stellar surface.

\subsection{Interstellar Scattering}

\subsubsection{Scintillation Physics}

Density inhomogeneities in the dilute, turbulent plasma of the ISM scatter the radio emission, leading to multipath propagation. At 760 MHz, the Vela pulsar is deep within the regime of `strong' scattering, for which the Fresnel scale is much larger than the diffractive scale. 
Furthermore, the timescale for evolution of the diffraction pattern at the observer is several seconds, so each scintillation element is effectively static for the duration of a single pulse (${\sim}5\mathrm{\ ms}$). The scattering then acts to convolve the pulsar signal with a stochastic propagation kernel, which reflects the scintillation and temporal broadening \citep{Hankins_Coherent,Williamson,Intermittent_Noiselike_Emission}.

The spatial extent of the emission region influences the scintillation. A pair of dimensionless parameters, $\left\{\gamma_{\mathrm{s},1},\gamma_{\mathrm{s},2}\right\}$, quantify the emission region and give its squared characteristic size in an orthogonal, transverse basis $\{ \hat{\textbf{x}}_1,\hat{\textbf{x}}_2 \}$, in units of the magnified diffractive scale \citep{ISO}:
\begin{align}
\label{eq::gamma_s}
\gamma_{\mathrm{s},i} = \left( \frac{D}{R} \frac{\sigma_i}{\frac{1}{2\pi}\frac{\lambda}{\theta_i} } \right)^2\!.
\end{align}
Here, $D$ is the characteristic observer-scatterer distance, $R$ is the characteristic source-scatterer distance, $\lambda$ is the observing wavelength, $\theta_i$ is the angular size of the scattering disk along $\hat{x}_i$, and $\sigma_i$ is the standard deviation of the distribution of source intensity along $\hat{x}_i$. 
This representation accommodates anisotropy of both the source emission and the distribution of scattering material. 

Nevertheless, the effects of anisotropy in the emission and scattering only manifest at quadratic order in $\gamma_{\mathrm{s}}$; thus, we use $\gamma_{\mathrm{s}} \equiv (\gamma_{\mathrm{s},1} + \gamma_{\mathrm{s},2})/2$ to avoid parameter degeneracy. Likewise, we state our results in terms of the standard deviation $\sigma_{\mathrm{c}} \equiv \sigma_1 = \sigma_2$ of a circular Gaussian intensity profile. The FWHM of such a profile is $2\sqrt{\ln 4} \sigma_{\mathrm{c}} \approx 2.35\sigma_{\mathrm{c}}$.

\subsubsection{Interpretation of the Inferred Emission Size}
The emission size that we infer from $\gamma_{\mathrm{s}}$ reflects the full extent of the emission region. Explicitly, we observe the spatial standard deviation of the emitting region, weighted by integrated flux density \citepalias{IPDF}. Thus, even if the instantaneous emission is pointlike, if the received radiation arises from spatially offset emission sites then $\gamma_{\mathrm{s}} > 0$. We derive the effective emission size of beamed, dipolar emission in \S\ref{sec::Interpretation}.

\subsubsection{Scattering Geometry of the Vela Pulsar}
\label{sec::kmConversion}

The conversion of $\gamma_{\mathrm{s}}$ to a physical size at the pulsar requires measurements of the location, scale, and nature of the scattering.  
Using VLBI parallax, \citet{Dodson_2003} estimated a distance of $287^{+19}_{-17}$ pc to the Vela pulsar. Combined with the angular and temporal broadening of the pulsar, this distance identifies the characteristic location of the scattering material \citep{Gwinn_Bartel_Cordes}. 
\citet{Gwinn97} measured the angular broadening to be ($3.3\pm0.2$ mas)$\times$($2.0\pm 0.1$ mas) (FWHM) at 2.8 GHz. Combined with the temporal broadening, as calculated via the decorrelation bandwidth $\Delta \nu_{\mathrm{d}} = 66\mathrm{\ kHz}$ at $2.3\mathrm{\ GHz}$, these measurements suggest a characteristic fractional distance of $\frac{D}{D+R} \approx 0.72$ to the scattering material and an effective magnification of $D/R\approx 2.9$ \citep{Gwinn00}. 
Thus, if the source emits a circular Gaussian intensity profile, then $\sigma_{\mathrm{c}} \approx \left( \frac{\nu_{\mathrm{MHz}}}{ 800\mathrm{\ MHz}} \right) \times 440 \sqrt{\gamma_{\mathrm{s}}} \mathrm{\ km}$. 

Different models of the scattering medium or emission geometry will update this conversion but will not affect our model PDFs or measurements of $\gamma_{\mathrm{s}}$. For example, we quantify the effect of the phase structure function of the scattering material in \S\ref{sec::PSF_Effects}.

\section{Observation and Data Reduction}
\label{sec::Obs}

\subsection{Observation and Baseband Recording}
\label{ssec::Obs_BBRec}

We observed the Vela pulsar for one hour on 22 October 2010 and one hour on 25 October 2010 using the Robert C.\ Byrd Green Bank Telescope (GBT) and the recently built Green Bank Ultimate Pulsar Processing Instrument ({\sc{guppi}}). For flux calibration, we also observed 3C\ 190 (0758+143) for approximately 20 minutes prior to each pulsar observation.

We used {\sc{guppi}} in 8-bit, 200 MHz mode, and our observation spanned $723.125{-}923.125$ MHz. 
Rather than utilizing the standard real-time dedispersion of {\sc{guppi}}, we recorded the channelized complex baseband voltages directly to disk for offline processing. 
The collected data constitute approximately 8 TB.

\subsection{Coherent Dedispersion and Formation of Spectra}

We processed the baseband data using the digital signal processing software library {\sc{dspsr}} \citep{DSPSR_ref}, which first performed coherent dedispersion and then formed dynamic spectra via a \textit{phase-locked filterbank}. This technique constructs each spectrum from a fixed-length time-series beginning at the corresponding pulse phase, rather than windowing the pulse into a fixed number of temporal phase bins. Our spectra had 262{,}144 channels per 25 MHz, corresponding to 95.4 Hz resolution. The ${\sim} 10.5\mathrm{\ ms}$ spectral accumulation time easily contained all of the power in one pulse. 

We exported the coherently-dedispersed spectra from {\sc{dspsr}} into {\sc{psrfits}} format and then performed flux calibration for each polarization using the software suite {\sc{psrchive}} \citep{PSRCHIVE_ref}. 
Subsequent processing utilized the {\sc{psrchive}} libraries, which we integrated into our custom software. 

Although reversing the Faraday rotation is straightforward, we chose to analyze the measured linear polarization streams without modification in order to avoid introducing possible artifacts from calibration errors and mixing. Consistency of our results between the two polarizations is, therefore, a meaningful and encouraging indication that our inferences reflect properties of the pulsar rather than instrumental limitations.

\subsection{Data Excision and Bandpass Correction}

\subsubsection{RFI Excision}
We iteratively excised RFI using a median smoothed difference criterion, similar to that implemented in {\sc{psrchive}}. Specifically, we compared the estimated background noise in each frequency channel with the noise in its surrounding $250\mathrm{\ kHz}$ sub-band; we rejected channels that differed from the sub-band median by more than three times the sub-band standard deviation. Such rejections typically constituted a tiny fraction of the data (${\sim} 0.3\%$, as expected from the $3\sigma$ cut). Also, to reject impulsive RFI, we dropped all pulses for which the off-pulse mean differed from a three-minute moving average of the off-pulse region by more than $5\%$.

\subsubsection{Background Noise and Bandpass Shape Estimation}
{\sc{guppi}} divided the observed 200 MHz frequency band into 32 filterbank sub-bands, each of 6.25 MHz. Because of spectral roll-off, we only analyzed the central 3.2 MHz of each filterbank sub-band; each such segment was analyzed independently.

We estimated both the background noise level and the bandpass shape by analyzing the off-pulse spectra. The off-pulse region excluded the $10.5$ ms on-pulse spectra, as well as a frequency-dependent phase range of width $0.15$ that contained negatively-dedispersed power leaked from the interleaved sampling. 

Next, we normalized the bandpass. We determined the bandpass shape for each three-minute block using a $(\mathrm{3\ minute})\! \times \! (\mathrm{100\ kHz})$ moving average of the off-pulse data, or ${\sim} 2\!\times\! 10^6$ samples for each spectral channel. The bandpass variations across each 3.2 MHz sub-band were typically a few percent; thus, they contributed minimally to the observed statistics because of their small variance.

\subsubsection{Rejection of Pulses with Quantization Saturation}
Because the Vela pulsar is both bright and variable, particularly when coupled with scintillation, some of the strongest pulses saturated our quantization thresholds. 
Moreover, because the quantization levels were fixed across the band, the strong linear polarization of the pulsar (combined with Faraday rotation) and the receiver bandpass shape caused the prevalence of saturation to vary substantially across the 200 MHz observing bandwidth. 

Accordingly, we implemented a five-level `guard zone' at the positive and negative thresholds (of 256 total levels); we excluded any pulse that included a quantized value $|x|>122$. This criterion rejected $\gtwid 75\%$ of pulses for the stronger linear polarization near $850\mathrm{\ MHz}$, but rejected $\lesssim 1\%$ of pulses, regardless of polarization, at frequencies $\ltwid 775\mathrm{\ MHz}$. To avoid a selection bias against strong pulses, we preferentially draw later examples from the lower-frequency sub-bands.

\subsection{Observed PDF of Intensity}
\subsubsection{Histogram Formation}
We constructed a histogram of measured intensities for every $(1\mathrm{\ hour})\! \times\! (3.2 \mathrm{\ MHz})$ block of on-pulse data (${\sim} 10^9$ spectral samples). Histogram bins had uniform widths of $0.05$, in units where the mean background intensity was unity because of the bandpass normalization. 
We then normalized each histogram by the total number of samples and the bin width; we will refer to this normalized histogram as the `observed PDF' of intensity.

\subsubsection{Effects of Finite Histogram Resolution}
\label{sec::Histogram_Resolution}
In order to compare a theoretical PDF $P(I)$ with an observed PDF, we must account for the finite bin width of the histogram. Specifically, the theoretical (normalized) histogram value $H(I;\Delta I)$ for a bin of width $\Delta I$ centered on $I$ is given by
\begin{align}
H(I;\Delta I) = \frac{1}{\Delta I}\int_{I - \Delta I/2}^{I + \Delta I/2} P(I') dI'. 
\end{align}
Expanding $P(I')$ about $I$ and replacing the second derivative with its discrete representation gives 
\begin{align}
H(I;\Delta I) &\approx P(I) + \frac{P(I + \Delta I) - 2 P(I) + P(I - \Delta I)}{24}.
\end{align}
This approximation corrects for the finite bin width to quadratic order in $\Delta I$, while only requiring calculation of the theoretical PDF at the histogram bin centers. 
For our data, the next order correction, $\Delta I^4 P^{(4)}(I)/1920$, was well below our Poisson noise. 
The effects of the finite bin width were small when $N=1$ (typically $\ltwid0.005\%$ of the PDF amplitude), but they became appreciable for larger $N$.

\subsection{Model PDF of Intensity}

We independently constructed observed PDFs for each scalar electric field (i.e.\ each linear polarization) and with various degrees of temporal averaging $N$, bandwidth, and pulse selection cuts. In all of these cases, the mean source and background intensities of the pulses parametrized the point-source model PDF of intensity; 
this parametrization accounted for the intrinsic pulse-to-pulse variability as well as the quasistatic evolution of receiver gain and noise levels. 
Because this treatment incorporated the observed set of pulse amplitudes rather than appealing to assumptions about their distribution, our models were robust to aribitrary pulse-to-pulse variations, including strongly correlated behavior such as nulling or heavy-tailed distributions of pulse amplitude, as are observed for some pulsars \citep{Crab_GPPL,Vela_lognormal}.

\subsubsection{Model PDF Parameter Estimation}

We estimated the single-pulse source and background amplitudes using the mean of all the on-pulse or off-pulse channels in the analyzed sub-band that had not been flagged for RFI. For a $3.2\mathrm{\ MHz}$ sub-band, this estimate included ${\sim} 30{,}000$ samples for the on-pulse region and ${\sim} 200{,}000$ samples for the off-pulse region of each pulse. This average effectively isolated the intrinsic pulsed power from scintillation because each averaged sub-band contained many independent scintillation elements ($\Delta \nu_{\mathrm{d}} \sim 1\mathrm{\ kHz}$). However, these estimates were subject to both bias and noise, and we determine the effects of such errors in \S\ref{sec::Errors_and_Bias}.

\subsubsection{Construction of the Model PDF}
We now describe the basis and formation of our model PDFs of intensity. 
First, consider $N$ consecutive pulses with gated source intensities $A_1 I_{\mathrm{s}},\dots,A_N I_{\mathrm{s}}$ and mean background intensity $I_{\mathrm{n}}$. After averaging the $N$ spectra formed from these pulses, the PDF of intensity is \citepalias[Eq.\ 3]{IPDF}
\begin{align}
\label{I_PDF}
P(I;N) = &N \int_0^\infty dG \ \frac{e^{-G}}{ \left(I_{\mathrm{s}} G\right)^{N-1} } \\
\nonumber & \qquad {} \times \sum_{j=1}^N  \frac{\left(A_j I_{\mathrm{s}} G  + I_{\mathrm{n}} \right)^{N-2} }{\prod\limits_{\substack{\ell = 1\\ \ell\neq j}}^N  \left(A_j - A_\ell \right)} e^{-\frac{N I}{\left(A_j I_{\mathrm{s}} G + I_{\mathrm{n}} \right)}}.
\end{align}
The integral over $G$ arises because we assume that each spectrum effectively explores the full scintillation ensemble.

We calculated this theoretical PDF for every sequence of $N$ pulses that contributed to an observed PDF, using the estimated model parameters $\{A_j I_{\mathrm{s}},I_n\}$. We then averaged these distributions to yield the `model PDF' of intensity. Note that the model PDF required no fitted parameters and was independently constructed for each observed PDF.

\subsection{Expected Residual Structure}
The PDF residual $\mathcal{R}(I)$ (i.e.\ the observed minus model PDF) contains signatures of extended emission size, decorrelation of the scintillation within the averaging time, and errors in the model parameters $\{ A_j I_{\mathrm{s}}, I_{\mathrm{n}} \}$. These perturbations have nearly identical shape, so a single fitted parameter quantifies their combined effect. We can break this degeneracy by varying the reduction parameters such as $N$ and the analyzed bandwidth. 
For example, the temporal decorrelation of the scintillation pattern does not contribute when $N=1$. 

We approximate the expected cumulative residual arising from the emission size and the temporal decorrelation as \citepalias[Eq.\ 9, 10, 18]{IPDF}
\begin{align}
\label{eq::Residual}
\mathcal{R}(I) &\approx - \left[2\gamma_{\mathrm{s}} + \frac{\sum_{i<j}\left(1 - \Gamma_{ij} \right)}{N(N+1)} \right] \\
\nonumber & \qquad {} \times \int_0^\infty dG \ G e^{-G} \frac{\partial^2 P(I;N|G)}{\partial G^2}.
\end{align}
This general form expresses the influence of the temporal autocorrelation function of the scintillation pattern:
\begin{align}
\Gamma_{ij} \equiv \frac{\langle \left[I(t_i, f) - I_{\mathrm{n}}(t_i, f)\right] \left[ I(t_j,f) - I_{\mathrm{n}}(t_j,f)\right] \rangle}{\langle I(t,f) - I_{\mathrm{n}}(t,f) \rangle^2} - 1. 
\end{align}
We will quantify alternative sources of residual structure, such as parameter errors, in terms of their incurred bias on $\gamma_{\mathrm{s}}$. As was required for the model PDF, this model residual was calculated for every sequence of $N$ pulses that contributed to the corresponding observed PDF and subsequently averaged.

If the $N$ averaged intensities are drawn from consecutive pulses, then the scale $\mathcal{R}_0$ of the residual structure can be approximated as
\begin{align}
\label{eq::Residual_N}
\mathcal{R}_0 \equiv \gamma_{\mathrm{s}} + \frac{\sum_{i<j}\left(1 - \Gamma_{ij} \right)}{4 N(N+1)} \approx \gamma_{\mathrm{s}} + \frac{1}{24} \frac{N(N-1)}{\Delta \tau_{\mathrm{d}}^2 },
\end{align}
where $\Delta \tau_{\mathrm{d}} \gg N$ is the decorrelation timescale in pulse periods, and we have assumed the standard square-law autocorrelation $\Gamma_{ij} = e^{-\left[ (i-j) / \Delta \tau_{\mathrm{d}} \right]^2}$.
Thus, the relative influence of temporal decorrelation to emission size is approximately quadratic in $N$.

\subsection{Fitting the Residual Structure}

For each observed PDF, we calculated the corresponding model PDF and residual shape. We then performed a weighted least-squares fit to the measured residual after correcting for the finite histogram resolution; we assigned weights by assuming that the histogram errors reflected Poisson noise. The only fitted parameter was the amplitude scale $\mathcal{R}_0$ of the residual. In some cases, we performed multiple reductions on the same set of pulses by varying the averaging ($N$) or the lag between the averaged pulses ($\Delta \tau$); we then fit all these reductions simultaneously to the pair of parameters $\{ \gamma_{\mathrm{s}},\Delta t_{\mathrm{d}} \}$ by assuming a square-law autocorrelation structure.

\section{Analysis}
\label{sec::Analysis}

We now describe our analysis and subsequent inference, in successively tighter pulse cuts. We began by analyzing statistics of each polarization, when pulses were all analyzed concurrently (\S\ref{sec:All_Pulses}). Then, we analyzed PDFs formed by averaging pairs of intensities with varying temporal lag, to demonstrate that our description of the effects of temporal decorrelation is correct (\S\ref{sec::Neq2}). Next, we segregated pulses into quantiles by pulse strength to investigate the possible covariance between the size of the emission region and pulse shape (\S\ref{sec::Quantiles}). Finally, we examined the statistics of each individual pulse (\S\ref{sec::sps}).

We did not detect the effects of a finite emission region in any of these cases, regardless of frequency or polarization. Thus, our chosen examples are merely those that yield the tightest constraints on $\gamma_{\mathrm{s}}$. For this purpose, we favor one filterbank sub-band centered on 757.5 MHz, for which the polarization position angle of the pulsar was nearly aligned with that of the receiver, the pulsar rarely saturated the quantization thresholds, the leaked power from the interleaved sampling was far off-pulse, and there was no evidence for substantial RFI contamination. To achieve our tightest limit on the size of the emission region, we jointly analyze this sub-band with the adjacent sub-bands. 
The full frequency evolution of our observations is a powerful probe of the scattering, on the other hand, as we analyze separately \citep{GUPPI_Scattering}.

\subsection{All Pulses}
\label{sec:All_Pulses}
To utilize the full statistical content of our observations, we first analyzed all pulses concurrently, in groups only determined by frequency and polarization. 
We constructed each observed PDF from a $(1\mathrm{\ hour})\! \times\! (3.2 \mathrm{\ MHz})$ on-pulse block of data: ${\sim}10^9$ spectral samples and ${\sim}40{,}000$ pulses. The scintillation pattern evolved substantially over both the span of time and frequency; each observed PDF contained samples from ${\sim}10^6$ independent scintillation elements. 

\begin{figure}[t]
\includegraphics*[width=0.478\textwidth]{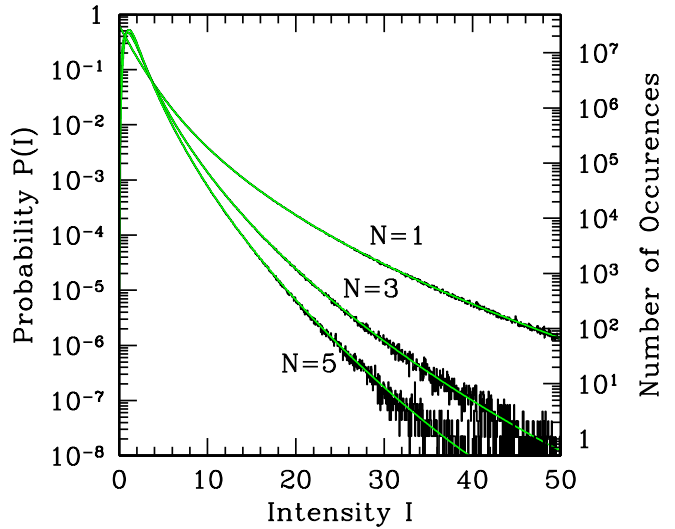}
\caption{
Observed and model PDFs of intensity for three degrees of averaging: $N=1$, 3, and 5. There are no fitted parameters. The mean intensities of the source and background are $I_{\mathrm{s}}\langle A \rangle = 0.87$ and $\langle I_{\mathrm{n}} \rangle = 1.00$, respectively; the plot therefore extends to ${\sim} 27$ times the mean signal. 
}
\label{fig_IPDF}
\end{figure}

Figure \ref{fig_IPDF} demonstrates one example of the comparison of our model and observed PDFs of intensity; the plotted data correspond to the stronger linear polarization for the 3.2 MHz sub-band centered on 757.5 MHz. 
The agreement extends over the nearly eight decades spanned. 
The tail of the distribution arises from the combined effects of intrinsic variability, noise, and scintillation; the success of the model indicates that it correctly accounts for each effect.

Our models also correctly characterize the effects of temporal averaging, $N$. The decorrelation timescale in pulse periods is $\Delta \tau_{\mathrm{d}} \sim 50$, so the correction for temporal decorrelation is quite small for the plotted reductions: $N{=}1$, 3, and 5 (see Eq.\ \ref{eq::Residual_N}).

For $N=1$, the residual structure in the stronger polarization has an amplitude of ${\ltwid} 10^{-4}$, which is comparable to the Poisson noise in the histogram. A weighted linear least-squares fit to the lowest 200 samples ($I < 10$) in the observed PDF estimates $\gamma_{\mathrm{s}} = (2.0 \pm 1.0){\times} 10^{-4}$ with reduced chi-squared $\chi_{\mathrm{red}}^2 = 1.13$; the standard deviation of $\chi^2_{\mathrm{red}}$ is $\sqrt{2/200}=0.1$. Furthermore, because intrinsic modulation correlates noise in adjacent spectral channels, Poisson weights underestimate the histogram errors, so $\chi^2 \!\gtwid\! 1$ is expected.  
Thus, a conservative upper-bound on the effects of size from this single sub-band is $\gamma_{\mathrm{s}} \ltwid 5\times10^{-4}$, which is equivalent to $\sigma_{\mathrm{c}} \approx 8\mathrm{\ km}$ at the pulsar with the conversion of \S\ref{sec::kmConversion}.

For the adjacent sub-bands (centered on 751.25 MHz and 763.75 MHz), the fits give $\gamma_{\mathrm{s}} = (-2.2 \pm 1.1) {\times} 10^{-4}$ and $\gamma_{\mathrm{s}} = (-1.2 \pm 1.0) {\times} 10^{-4}$. Combining these results refines the 3$\sigma$ size upper-bound to approximately 4 km at the pulsar. Narrower sub-bands give similar results, with larger errors because of the increased Poisson noise and a larger bias from the finite bandwidth (see \S\ref{sec::Parameter_Noise}). In all cases, our results suggest a characteristic size of ${\ltwid} 10\mathrm{\ km}$ for the emission region.

\subsection{$N=2$ with Temporal Lag $\Delta \tau$}
\label{sec::Neq2}
We next examined effects dominated by temporal decorrelation of the scintillation pattern. We constructed PDFs with $N=2$ that averaged pairs of pulses separated by $\Delta \tau$ pulse periods;
the amplitude of the residual structure (Eq.\ \ref{eq::Residual}) then depends roughly quadratically on $\Delta \tau$. We approximated the autocorrelation function as Gaussian, $\Gamma(\Delta \tau) = e^{-\left(\Delta \tau/\Delta \tau_{\mathrm{d}} \right)^2}\!$, to simultaneously fit all lags to the pair of parameters, $\{ \gamma_{\mathrm{s}},\Delta \tau_{\mathrm{d}}\}$.

\begin{figure}[t]
\includegraphics*[width=0.478\textwidth]{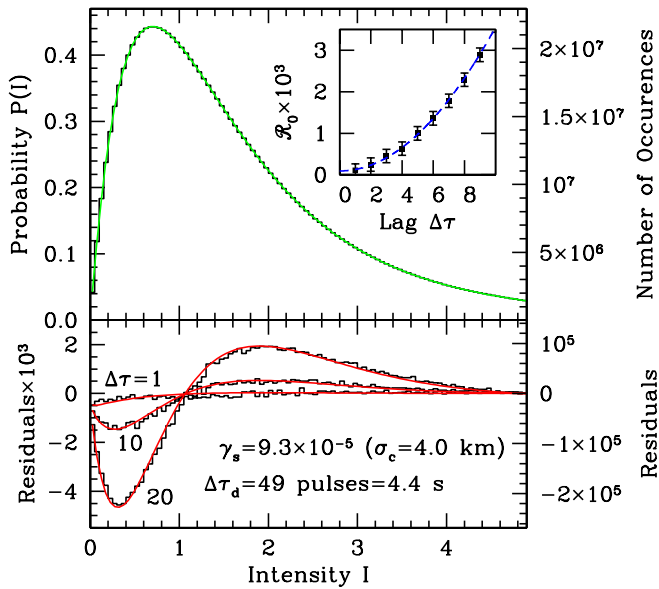}
\caption{
Observed and model PDFs of intensity with their corresponding residual structure; $N=2$. Three cases are plotted, corresponding to averaged pairs of intensities that are separated by $\Delta \tau = 1$, 10, and 20 pulses. The residuals are concurrently fit to the decorrelation timescale $\Delta t_{\mathrm{d}}$ and dimensionless size $\gamma_{\mathrm{s}}$. The inset shows the fitted residual scale $\mathcal{R}_0$ with $3\sigma$ errorbars as a function of $\Delta \tau$, along with the best-fitting prediction (see Eqs.\ \ref{eq::Residual} and \ref{eq::Residual_N}). 
}
\label{fig_IPDF_decorr}
\end{figure}

The residuals match their expectations quite well; Figure \ref{fig_IPDF_decorr} shows one example, again from the stronger linear polarization for the 3.2 MHz sub-band centered on 757.5 MHz. This excellent agreement demonstrates the quality of our analytical representation of decorrelation of the scintillation pattern within the averaging time. The fits estimate $\gamma_{\mathrm{s}} = (9.3 \pm 5.4){\times}10^{-5}$ giving a $3\sigma$ upper-bound on the size of the emission region as ${\sim}7\mathrm{\ km}$.

\subsection{Quantiles}
\label{sec::Quantiles}
Individual pulse profiles for the Vela pulsar show marked dependence on their strength \citep{Krishnamohan}. To examine covariance between pulse strength and emission size, we segregated the pulses into quantiles by their mean flux density and then analyzed each subset independently. 

This approach has several utilities. Because the resolving power improves with the SNR, a subset of strong pulses can maximize the potential resolution. Furthermore, many instrumental biases vary differently with pulse strength than their effects on the inferred size do (see \S\ref{sec::Parameter_Bias}); the variation of $\gamma_{\mathrm{s}}$ between quantile subsets can therefore suggest the presence of such a bias.

\begin{figure}[t]
\includegraphics*[width=0.478\textwidth]{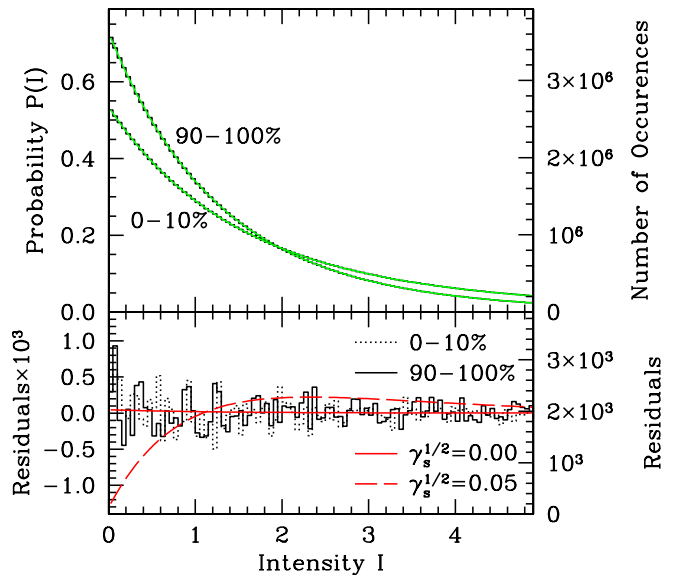}
\caption{
Observed and model PDFs of intensity for the subsets of pulses in the top and bottom decile by pulse intensity. Theoretical residual curves are plotted for a point source (the slight variation is from finite histogram resolution) and for a source that extends over 5\% of the magnified diffractive scale (i.e.\ $\sigma_{\mathrm{c}}\approx 20\mathrm{\ km}$); the latter is clearly inconsistent with the observed statistics.
}
\label{fig::decile}
\end{figure}

We established deciles by pulse strength. Figure \ref{fig::decile} shows the PDFs and their residual structure for the top and bottom decile; the mean gated signal-to-noise $\langle A_j I_{\mathrm{s}}/I_{\mathrm{n}} \rangle$ in these deciles is $1.39$ and $0.48$, respectively. The two residuals are comparable, and both suggest an emission size that is substantially smaller than the magnified diffractive scale. The limit is most stringent for the pulses in the top decile because of their higher SNR: $\gamma_{\mathrm{s}} = \left(1.1 \pm 1.8 \right) \times 10^{-4}$, or an 11 km upper-bound on the characteristic emission size at a 3$\sigma$ confidence level. For the bottom decile, we obtain $\gamma_{\mathrm{s}} = \left( -7.9 \pm 5.5 \right) \times 10^{-4}$, which gives a 12 km upper-bound at the same confidence.

\subsection{Single-Pulse Statistics}
\label{sec::sps}
Lastly, we examined the statistics for individual pulses. For each pulse, our model requires only two parameters to describe the spectral statistics: the mean intensities of the background ($I_{\mathrm{n}}$) and gated signal ($A I_{\mathrm{s}}$). Poisson noise generally limits each estimate of emission size to a sizeable fraction of the magnified diffractive scale; the standard error $\delta \gamma_{\mathrm{s}}$ in the fitted $\gamma_{\mathrm{s}}$ is well-approximated by \citepalias[Eq.\ 26]{IPDF}
\begin{align}
\label{eq::Size_Error}
\delta \gamma_{\mathrm{s}} \approx \frac{1.3}{\sqrt{N_{\mathrm{tot}}} S} \exp\left[0.4 + 0.18 \left(\ln S\right)^2 + 0.01 \left(\ln S \right)^3 \right].
\end{align}
Here, $N_{\mathrm{tot}}$ is the total number of points used to estimate the observed PDF, and $S\equiv A_j I_{\mathrm{s}} / I_{\mathrm{n}}$ is the gated signal-to-noise ratio . 
Thus, for a strong pulse ($S \sim 2$), $\gamma_{\mathrm{s}}$ can be determined to within $\pm 0.006$; whereas, for a weak pulse ($S=0.1$), $\gamma_{\mathrm{s}}$ can only be established to within $\pm 0.2$. 
Interestingly, for observations of highly variable pulsars, the resolution afforded by strong single pulses may exceed that of the entire pulse ensemble analyzed jointly.

\begin{figure}[t]
\includegraphics*[width=0.478\textwidth]{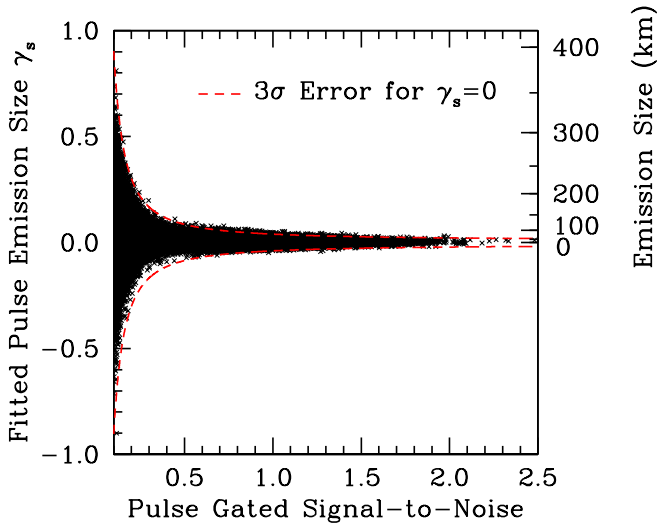}
\caption{
Inferred emission sizes of individual pulses as a function of their gated signal-to-noise $S$; both linear polarizations are plotted. Because $S$ determines the standard error for each measurement to excellent accuracy, we omit errorbars and instead show the expected $\pm 3\sigma$ errors about $\gamma_{\mathrm{s}}=0$. We do not obtain a statistically significant detection of emission size for any pulse.
}
\label{fig::spfit}
\end{figure}

Figure \ref{fig::spfit} shows the fitted sizes for ${\sim}30{,}000$ individual pulses as a function of their gated signal-to-noise; values from both linear polarizations are plotted. This sample of pulses included several `micro-giant' pulses, as well as tens of pulses with the late `bump' component identified by \citet{Johnston_Vela}. We measured no confident detection of emission size effects; the inferred values of $\gamma_{\mathrm{s}}$ are consistent with the expected standard errors for observations of a source with point-like emission, as given by Eq.\ \ref{eq::Size_Error}.

\section{Sources of Bias in the Estimated Emission Size}
\label{sec::Errors_and_Bias}
Many practical limitations can bias the inferred emission size \citepalias[\S4.4]{IPDF}. For instance, the model parameters $A_j$ and $I_{\mathrm{n}}$ are random variates estimated via finite averages of on-pulse and off-pulse spectra; they are therefore subject to noise, which is unbiased, and instrumental limitations such as quantization, which can introduce bias. Both of these types of errors can mimic the effects of an extended emission region. Here, we estimate the bias to $\gamma_{\mathrm{s}}$ resulting from each of these effects. We also account for uncertainties in the detailed distribution and nature of the scattering material.

\subsection{Effects of Noise in Model Parameters}
\label{sec::Parameter_Noise}
Noise in the model parameters introduces spurious modulation. 
Subsequent fits to the spectral statistics will then exhibit a positive bias in $\gamma_{\mathrm{s}}$ to quench this excess modulation. 
Fortunately, we can readily characterize the expected noise $\delta A_j$ and $\delta I_{\mathrm{n}}$ and can therefore anticipate the influence of such a bias.

For example, the noise $\delta A_j$ in the source amplitudes is dominated by the finite sampling of the scintillation pattern for a single spectrum; the corresponding bias is $\gamma_{\mathrm{s}} \rightarrow \gamma_{\mathrm{s}} + 1/(4 N_{\mathrm{scint}})$ \citepalias[\S4.4.2]{IPDF}, where $N_{\mathrm{scint}}$ is the number of averaged scintillation elements used to estimate $A_j$. 

On the other hand, because the background noise is white and approximately stationary, $\delta I_{\mathrm{n}}$ only depends on the number of samples $N_{\mathrm{b}}$ that are averaged to estimate $I_{\mathrm{n}}$. However, the incurred bias to $\gamma_{\mathrm{s}}$ also depends on the gated signal-to-noise ratio $S$: $\gamma_{\mathrm{s}} \rightarrow \gamma_{\mathrm{s}} + 1/(4 S^2 N_{\mathrm{b}})$.

For each 3.2 MHz sub-band, we had $N_{\mathrm{scint}} \sim (3.2 \mathrm{\ MHz})/\Delta \nu_d \approx 3{,}000$, and $N_{\mathrm{b}} \sim 10^5$. Both biases to $\gamma_{\mathrm{s}}$ were therefore within our quoted confidence. Furthermore, each bias scales as the reciprocal of the analyzed bandwidth. We observed no evolution in the inferred values of $\gamma_{\mathrm{s}}$ for 1.6 MHz and 0.8 MHz sub-bands and thus confidently established that this class of parameters error did not substantially influence our results.

\subsection{Effects of Bias in Model Parameters}
\label{sec::Parameter_Bias}
Instrumental limitations can also lead to errors in the model parameters. For example, interleaved sampling results in a negatively-dedispersed pulse `echo'; for highly dispersed pulsars, such as the Vela pulsar, this process can therefore leak pulsed power into the off-pulse region, leading to a positive bias in the estimated background noise: $I_{\mathrm{n}} \rightarrow I_{\mathrm{n}} + \delta I_{\mathrm{n}}$. Alternatively, quantization introduces signal-dependent noise \citep{Jenet_Anderson,Gwinn_Quant_Noise}; the off-pulse region then underestimates the effective on-pulse noise. 

Of course, any bias in the estimated background noise incurs a corresponding bias in the estimated source amplitude, and vice versa: $\delta I_{\mathrm{n}} = -\delta A_j I_{\mathrm{s}}$. These errors then affect the inferred emission size as $\gamma_{\mathrm{s}} \rightarrow \gamma_{\mathrm{s}} -  \delta I_{\mathrm{n}}/(2 A I_{\mathrm{s}})$. Because the dependence of emission size with pulse strength may vary differently than the noise bias, segregating pulses by their gated signal-to-noise can identify the influence of this type of bias. Because we did not observe variation in $\gamma_{\mathrm{s}}$ between decile subsets, this type of parameter error is unlikely to have significantly corrupted our emission size inference.

We further tested for the presence of a noise bias by comparing the estimates of $I_{\mathrm{n}}$ as a function of pulse strength. Leaked power, for instance, would have introduced a bias that was positively correlated with pulse strength. We detected a slight \emph{negative} trend in our estimates of $I_{\mathrm{n}}$ with pulse strength; the variation in $\langle I_{\mathrm{n}} \rangle$ between the top and bottom deciles (as shown in Figure \ref{fig::decile}) is approximately $0.15\%$. If this variation reflects a bias in the estimated noise, then the incurred contribution to $\gamma_{\mathrm{s}}$ is at approximately the $3\sigma$ errors we quote. However, the covariance between the fitted values of $\gamma_{\mathrm{s}}$ and $\langle I_{\mathrm{n}} \rangle$ is marginal, suggesting that this variation in $\langle I_{\mathrm{n}} \rangle$ is not responsible for any appreciable bias in the emission size inference. The variation may instead reflect noise in $I_{\mathrm{n}}$, which weakly anti-correlates with the gated signal-to-noise of the corresponding pulse. Then, for the full pulse ensemble, the error estimates of \S\ref{sec::Parameter_Noise} are still appropriate, although the segregation into quantiles will somewhat exaggerate this effect. 

Overall, these tests suggest that our 8-bit quantization scheme and high-quality backend ({\sc{guppi}}) effectively controlled any systematic errors in the model parameters. Remarkably, the determination of $\gamma_{\mathrm{s}}$ is then limited by the Poisson noise in the observed PDF of intensity (constructed from ${\sim}10^9$ samples). Because the transverse size at the source is proportional to $\sqrt{\gamma_{\mathrm{s}}}$, our resolving power scales as the fourth root of the observing duration and bandwidth.

\subsection{Effects of Modified Scattering Assumptions}
\label{sec::PSF_Effects}

Thus far, we have assumed thin-screen scattering with a square-law phase structure function, as is observationally motivated for the Vela pulsar \citep{Williamson,Komesaroff72,Roberts_Ables,Johnston_Scintillation,GUPPI_Scattering}. However, even with relaxed scattering assumptions, the same form of Eq.\ \ref{eq::gamma_s} still applies, with modified scaling.

For example, the translation to a general power-law structure function $D_\phi(\textbf{x}) \propto | \textbf{x} |^\alpha$ is given by \citep{GUPPI_Scattering}
\begin{align}
\gamma_{\mathrm{s},\alpha} = \left[ \frac{2^{3-\alpha}}{\alpha \Gamma\left( \frac{\alpha}{2} \right)} \gamma_{\mathrm{s}}  \right]^{2/\alpha}\!.
\end{align}
Hence, if the phase structure function is Kolmogorov ($\alpha = 5/3$), then the effects of a small emission size $\gamma_{\mathrm{s}} \ll 1$ will be somewhat more pronounced than for square-law scattering. The inferred transverse sizes will therefore be \emph{smaller} if we assume $\alpha = 5/3$ than those if we assume $\alpha =2$. Our given limits $\gamma_{\mathrm{s},\alpha=2} \ltwid 10^{-4}$ translate to $\gamma_{\mathrm{s},\alpha=5/3} \ltwid 2.3 \times 10^{-5}$, so this alternative scattering environment roughly halves our inferred limits.

The translation from $\gamma_{\mathrm{s}}$ to a dimensionful emission size also depends on the bulk scattering geometry (see Eq.\ \ref{eq::gamma_s}). In this case, the uncertainty in the magnification $D/R$ directly affects the inferred emission size. However, the excellent determination of both the temporal and angular broadening of the Vela pulsar lead to relatively small uncertainties for the magnification \citep{Gwinn00}, and thus the uncertainties for the inferred emission size are within the standard errors from Poisson noise.

\subsection{Effects of Proper Motion}
Because $\gamma_{\mathrm{s}}$ represents a characteristic size of the emission region, effects such as the proper motion of the pulsar can be important. However, the Vela pulsar has a transverse velocity of only ${\sim} 60$ km/s in its local environment \citep{Dodson_2003} -- quite slow relative to other young pulsars and the typical neutron star birth velocity \citep{Lyne_Lorimer}. Over our 10 ms accumulation time, the pulsar travels only $600\mathrm{\ m}$. Moreover, since size effects are weighted by pulse amplitude, the relevant timescale is the mean width of individual pulses, typically ${\sim} 2$ ms, contributing an effective displacement of ${\sim} 100\mathrm{\ m}$ -- much smaller than our present detection threshold.

\section{Interpretation}
\label{sec::Interpretation}

The simple, conventional picture of pulsar radio emission beamed along lines of a dipolar field provides contact between our measurements and those of traditional approaches to estimate the emission altitude $r_{\rm em}$. Specifically, this picture enables a translation between the effective transverse size of the emission region, $\sigma_{\rm c}$, and $r_{\rm em}$. Our measurements of $\sigma_{\rm c}$ imply tight limits on $r_{\rm em}$, in the context of this simple model, while obviating many assumptions of complementary techniques. 

In \S\ref{sec::Dipole_Model}, we review the basic geometric model for pulsar emission. 
Then, in \S\ref{sec::Altitude_Size}, we demonstrate how this model relates the transverse size of the emission region to the emission altitude.
Next, in \S\ref{sec::Alternate_Altitude}, we compare this technique and its result with traditional methods and their conclusions. Finally, in \S\ref{sec::Size_Compare}, we discuss possibilities for the discrepancy between our result and previous scintillation analyses of the Vela pulsar.

\subsection{Geometric Model for Pulsar Emission}
\label{sec::Dipole_Model}

A highly-simplified, conventional model describes the geometry of pulsar emission at centimeter and longer wavelengths. 
\citet{DRH} provide an overview of this model, with a discussion of its assumptions and parameters. 
We now outline the salient features for completeness. 
The model assumptions include a dipolar magnetic field, 
radio-wave emission tangent to field lines, 
and a single emission altitude for identifiable features of the average-pulse profile. 
This model follows from theoretical predictions of an ultra-relativistic electron-positron wind, streaming along the open dipolar field lines in the co-rotating frame \citep{Goldreich_Julian,Ruderman_Sutherland,Melrose_2004}, so we assume that the emission is beamed tangent to the field lines into an infinitesimal solid angle. 
In this case, an observer receives radiation from a single point in the pulsar's magnetosphere at each pulse phase. 
The variations of the spatial displacement of this point with pulse phase then determine the apparent size of the emission region. 

Because we merely seek a rough correspondence between the size and altitude of the emission region, we ignore 
effects of relativistic terms, modifications to the dipolar field structure, and magnetospheric ducting and reprocessing \citep{Blaskiewicz_91,Phillips_92,Kapoor_Shukre,Hibschman_01,Arons_Barnard,Lai_2001}.

A pair of parameters define the orientation of the pulsar: the inclination of the observer relative to the rotational axis ($\zeta$) and the impact angle of the magnetic pole with the line of sight ($\beta$). We also employ standard angular coordinates: the longitude or rotational phase of the pulsar ($\phi$) and the angle between the point of emission and the magnetic pole ($\theta_{\rm em}(\phi)$). For a fixed emission altitude $r_{\rm em}$, the spatial displacement of the point of emission $\vec D_{\rm em}(\phi)$ then follows from the dipolar geometry: $|\vec D_{\rm em}| \approx r_{\rm em} \tan^{-1}(\theta_{\rm em}/2)$. 
The cosine formula of spherical trigonometry relates $\theta_{\rm em}$ to pulse longitude $\phi$ and the orientation parameters $\{ \beta, \zeta \}$, and the well-known formula for the swing of position angle with $\phi$ gives the direction of $\vec D$ \citep{Komesaroff_1970}.
\citet[Figure 16.14]{PA} and \citet[Figure 3.4]{HPA}
provide useful figures and discussion.

The Vela pulsar benefits from several features that permit confident inferences of the orientation parameters. Fits to the X-ray emission from its pulsar wind nebula yield $\zeta\approx 63^{\circ}$ \citep{Ng_Romani_1,Ng_Romani_2}. In addition, the swing of the position angle of polarization across the pulse indicates that $\beta\approx -6.5^{\circ}$ \citep{RVM,Johnston_2005}. Thus, the Vela pulsar is a nearly-orthogonal rotator, with the line of sight close to the plane of rotation.
Note that the duty cycle of the Vela pulsar is quite small. The full angular extent of the emission is only $\rho \approx 16^\circ$ \citep{Johnston_Vela}, and the effective angular extent, expressed as the standard deviation in pulse phase weighted by flux density, is only $\rho_{\mathrm{eff}} = 3.0^{\circ}$, from our measurements.

\subsection{Emission Altitude from Effective Transverse Size}
\label{sec::Altitude_Size}
We now relate the effective emission size $\sigma_{\rm c}$ to $r_{\rm em}$, within the context of this geometrical model. A perfectly orthogonal rotator ($\beta=0$) provides an enlightening demonstration. In this case, the standard deviation of the transverse spatial displacement, weighted by flux-density, is $\sigma_{\rm eff} \approx \rho_{\rm eff} r_{\rm em} /3$. The factor of $1/3$ reflects that the angular difference between the tangent and radial directions of the dipolar field is roughly one third the colatitude of the tangent direction. The translation to an effective circular emission region $\sigma_{\rm c}$ then incurs an additional factor: $\sigma_{\rm eff} = \sqrt{2}\sigma_{\rm c}$. Applying our most stringent limits for the size of the emission region, $\sigma_{\rm c} < 4\mathrm{\ km}$, we obtain $r_{\rm em} < 3 \sqrt{2} \sigma_c/\rho_{\rm eff} \approx 320\ \mathrm{km}$. 

Applying the measured geometry for a nearly-orthogonal rotator, appropriately weighted by the individual pulse profiles for the stronger linear polarization at 757.5 MHz, we find $r_{\rm em}\approx 86 \sigma_{\rm c} <340\mathrm{\ km}$. Moreover, if we assume that the phase structure function of the scattering medium is Kolmogorov (see \S\ref{sec::PSF_Effects}), then we obtain $r_{\mathrm{em}} < 170\mathrm{\ km}$.

\subsection{Comparison with Previous Emission Altitude Estimates}
\label{sec::Alternate_Altitude}

Our results are comparable with the typical conclusions derived from pulse profile analysis and polarimetry. For example, \citet{Kijak_Gil_97} calculated pulsar emission heights by measuring pulse profile widths at 0.4 and 1.4 GHz and assuming that the emission spanned the open field-line region at a single altitude. For the Vela pulsar at 757.5 MHz, their model predicts $r_{\mathrm{em}} \approx 360\mathrm{\ km}$. 

More generally, the emission may arise from field lines that span only a portion of the open field-line region. Researchers typically characterize this possibility with the fractional colatitude of active field lines $f_{\theta} \approx \theta_{\rm max} \sqrt{R_{\rm LC}/r_{\rm em}}$.
The inferred emission altitude is quite sensitive to the assumed emission structure, scaling as ${\sim}f_\theta^{-2}$. However, with strict symmetry assumptions, $f_\theta$ can be inferred in pulsars with clearly delineated core and cone components \citep{GG2003,DRH}. Such measurements suggest that $f_{\theta}$ generally lies within the range $0.2{-}0.8$ for conal emission. 

Alternatively, the aberrational shift in position angle indicates the emission height \citep{Blaskiewicz_91,Phillips_92,Hibschman_01}. This shift is insensitive to $f_\theta$ and thus gives a robust $r_{\mathrm{em}}$ without constraining $f_\theta$. For the Vela pulsar, the position angle offset indicates $r_{\rm em} \sim 100\mathrm{\ km}$ \citep{Johnston_Vela}. 

A measurement of $\sigma_{\rm c}$ concurrently determines $f_\theta$ and $r_{\rm em}$, regardless of the profile morphology or the scatter-broadening. Thus, it can probe the emission structure at low frequencies, where both the size of the emitting region and the effects on the scintillation are most pronounced.

\subsection{Comparison with $1.66\mathrm{\ GHz}$ and $2.3\mathrm{\ GHz}$ Scintillation Studies}
\label{sec::Size_Compare}
The sizes that we infer for the emission region are substantially smaller than those found by the previous scintillation analyses discussed in \S\ref{sec::Previous}; however, our lack of resolution in pulse phase precludes a direct comparison. For example, if the emitting regions are largest when the pulsar is weak, then the effects of emission size will be diminished. Indeed, the phase-resolved analysis at $1.66\mathrm{\ GHz}$ suggests precisely this scenario. 
Furthermore, our observations span lower frequencies than previous analyses. Although the radius-to-frequency mapping \citep{Ruderman_Sutherland,Cordes_RFM} suggests that the emission height decreases with increasing frequency, the Vela pulsar possibly does not conform to this picture. 
Alternatively, the pulse profile shows considerable intrinsic evolution at 1 GHz, and the disparity between the measured emission size may indicate the introduction of a new, offset emission component at these higher frequencies. 
Pending observations should resolve this intriguing discrepancy.

\section{Summary}
\label{sec::Summary}

We have analyzed flux-density statistics of the Vela pulsar at 760 MHz. In particular, we examined the observed PDF of intensity in various cuts determined by frequency, polarization, and pulse strength. Based on single pulse estimates of the background noise and gated signal-to-noise, we constructed model PDFs in order to identify the signature arising from a spatially-extended emission region. 

The predicted form for point-source emission showed excellent agreement with the data ($\chi^2_{\mathrm{red}} {\approx} 1$), good to ${\sim}0.01\%$ over the low to moderate intensity regions of the PDF, in which the effects of an extended emission region are the most pronounced. This type of comparison requires no fitted parameters and arises from minimal assumptions about the characteristics of both the scattering (i.e.\ strong and slowly evolving) and source emission (i.e.\ amplitude-modulated noise with small spatial extent). Our inferences are therefore extremely robust. 

We did not detect any effects arising from an extended emission region. 
We achieved the tightest limit by jointly analyzing all pulses in three 3.2 MHz subbands, centered on 751.25, 757.5, and 763.75 MHz; 
this limit corresponds to a characteristic size of the emission region of $\sigma_{\mathrm{c}} \ltwid 4\mathrm{\ km}$ at the pulsar at $3\sigma$ confidence. 
This limit involved information about the nature and distribution of the scattering material. If the phase structure function of the scattering material is Kolmogorov, for instance, then the limit decreases to $\sigma_{\mathrm{c}} \ltwid 2\mathrm{\ km}$. 

The inferred size, $\sigma_{\mathrm{c}}$, reflects the spatial extent of all the emitting regions during a pulse and, therefore, also gives an upper-bound on the instantaneous emission size, which we assume is pointlike in our derivation of the emission altitude. 
Explicitly, $\sigma_{\rm c}$ corresponds to the standard deviation of a circular Gaussian distribution of intensity that yields equivalent scintillation statistics as those observed. Conversion to an emission altitude also involves the effective angular pulse width, weighted by flux density, which provides a natural mapping to lateral emission structure. We thereby obtained constraints on the emission altitude: $r_{\mathrm{em}} < 340\mathrm{\ km}$ (square-law) or $r_{\mathrm{em}} < 170\mathrm{\ km}$ (Kolmogorov). 

We also individually tested pulses for the signature of an extended emission region, again without a statistically significant detection. For the strongest pulses we obtained limits $\ltwid 50\mathrm{\ km}$, and for the majority of pulses we obtained limits $\ltwid 200 \mathrm{\ km}$. However, even for the strongest pulses, this bound only marginally constrained the emission to be generated within the light cylinder. Nevertheless, our results demonstrate the application and reliability of this technique.

Our upper bound on the size of the emission region improves previous estimates using scintillation statistics by over an order of magnitude. Without invoking symmetry constraints or assumptions about the distribution of emission within the open field-line region, we achieve estimates of the emission altitude that are comparable with alternative techniques that analyze profile widths and polarization characteristics. Thus, even for incredibly compact `core' emission, scintillation statistics now have the ability to refine and extend the information conveyed by bulk polarimetry. 

The excellent agreement between our data and models demonstrates that our technique can be similarly applied to other pulsars without fear of spurious inference. In particular, the quality of our data appears to have controlled the systematic uncertainties to be within the fundamental restriction set by Poisson noise. Thus, future observations, perhaps at lower frequencies or of other pulsars with favorable scattering geometries, may confidently measure a spatially-extended emission region and thereby greatly enrich our empirical understanding of pulsar radio emission.

\acknowledgments
We gratefully acknowledge the helpful efforts and guidance of Willem van Straten, particularly in writing the phase-locked filterbank of {\sc{dspsr}}, which was central to our analysis.  
We thank the U.S. National Science Foundation for financial support for this work (AST-1008865). The National Radio Astronomy Observatory is a facility of the National Science  Foundation operated under cooperative agreement by Associated Universities, Inc.

{\it Facilities:} \facility{GBT (GUPPI)}

\bibliography{Finite_Size_2_Short_References.bib}

\end{document}